# Ferromagnetic interlayer coupling in CrSBr crystals irradiated by ions

Fangchao Long[1,2], Mahdi Ghorbani-Asl[1], Kseniia Mosina[3], Yi Li[1,2], Kaiman Lin[1,4], Fabian Ganss[1], René Hübner[1], Zdenek Sofer[3], Florian Dirnberger[5*], Akashdeep Kamra[6], Arkady V. Krasheninnikov[1], Slawomir Prucnal[1], Manfred Helm[1,2], Shengqiang Zhou[1*]

[1]Helmholtz-Zentrum Dresden-Rossendorf, Institute of Ion Beam Physics and Materials Research, Bautzner Landstrasse 400, 01328 Dresden, Germany

[2]TU Dresden, 01062 Dresden, Germany

[3]Department of Inorganic Chemistry, University of Chemistry and Technology Prague, Technická 5, 166 28 Prague 6, Czech Republic

[4]University of Michigan-Shanghai Jiao Tong University Joint Institute, Shanghai Jiao Tong University, Shanghai, China

[5]Institute of Applied Physics and Würzburg-Dresden Cluster of Excellence ct.qmat, TU Dresden, 01062 Dresden, Germany

[6]Condensed Matter Physics Center (IFIMAC) and Departamento de Física Teórica de la Materia Condensada, Universidad Autónoma de Madrid, Madrid, Spain

- Corresponding authors: florian.dirnberger@tu-dresden.de, s.zhou@hzdr.de

## Abstract

Layered magnetic materials are becoming a major platform for future spin-based applications. Particularly the air-stable van der Waals compound CrSBr is attracting considerable interest due to its prominent magneto-transport and magneto-optical properties. In this work, we observe a transition from antiferromagnetic to ferromagnetic behavior in CrSBr crystals exposed to high-energy, non-magnetic ions. Already at moderate fluences, ion irradiation induces a remanent magnetization with hysteresis adapting to the easy-axis anisotropy of the pristine magnetic order up to a critical temperature of 110 K. Structure analysis of the irradiated crystals in conjunction with density functional theory calculations suggest that the displacement of constituent atoms due to collisions with ions and the formation of interstitials favors ferromagnetic order between the layers.

**Keywords:** 2D magnets, 2D semiconductor, Defects, Ion irradiation, Induced ferromagnetism

Van der Waals (vdW) magnets are attracting considerable interest from the scientific community. The ability to isolate single layers that can be reassembled into complex hetero-structures makes them particularly useful for spin-based technologies.[1] Two-dimensional (2D) ferromagnetism[2, 3] has first been found in CrI$_3$ and Cr$_2$Ge$_2$Te$_6$, but these materials are not stable under ambient conditions.[4, 5] For instance, CrI$_3$, a prototypical vdW magnet, deteriorates within minutes in air, which strongly hampers the fabrication and investigation of devices made from this compound. Hence, considerable efforts devoted to the search for new vdW magnetic crystals brought forth a number of air-stable antiferromagnetic (AFM) compounds, such as those from the *MPX*$_3$ family (*M* = Mn, Fe, Ni; *X* = S, Se)[6], and the layered magnetic semiconductor CrSBr.[7, 8] However, to exploit the full potential of vdW magnetic crystals for spin-based applications that require ferromagnetic (FM) coupling between magnetic moments, like magnetic memory devices, new experimental methods to control the magnetic coupling in vdW materials have to be developed.[9, 10]

Due to the pronounced correlations between magnons, photons, electrons, and phonons[7, 11-14], CrSBr is emerging as one of the most promising materials for such applications. Up to the Néel temperature of $T_N$ = 132 ± 1 K, the FM spin order within each layer is compensated by an AFM arrangement of the vdW layers in the out-of-plane direction (*c*-axis)[15], which is the reason for the vanishing net magnetization of pristine bulk crystals. First-principles calculations of magnetic moments localized on the Cr ions predict a magnetic order up to 160 K in a single layer.[7, 16] Second harmonic generation[17] and magneto-transport measurements[7, 18] determined the FM order of a single CrSBr layer to be around 150 K. Overall, bulk CrSBr is characterized by a relatively weak AFM coupling between the vdW layers, which is also apparent from the fact that a small field of 0.4 T applied along the easy-axis (*b*-axis) is sufficient to switch the magnetic order from AFM to FM.[11] The weak interaction between the layers makes CrSBr particularly susceptible to the modification of the magnetic order by external stimuli. Besides the recently demonstrated feasibility of strain and ligand substitution in altering the magnetic properties of CrSBr,[19-22] ion irradiation may thus be a viable tool for modifying the magnetic structure with additional benefits, such as local patterning.

Here, we report a change of the magnetic interlayer coupling in bulk and few-layer CrSBr crystals irradiated with non-magnetic ions. In the absence of an applied field and below a critical temperature of 110 K, the irradiated samples show sizeable spontaneous magnetization which adapts to the easy-axis anisotropy of the pristine magnetic order and marks the transition of the magnetic ground state of irradiated CrSBr crystals from AFM to FM. Raman spectroscopy evidences the ion-fluence-dependent softening of the crystal lattice that is indicative of the formation of a large number of crystallographic defects. In conjunction with structure analysis, first-principles calculations suggest that the displacement of the atoms, particularly Cr, into interstitial positions between the vdW layers favors FM interlayer coupling. Our study highlights the potential of ion irradiation for non-chemical engineering of magnetism in vdW crystals.

Millimeter-sized bulk CrSBr crystals are synthesized using a chemical vapor transport method (see SM, Methods). Micro-Raman spectroscopy is performed using a linearly polarized, continuous-wave, 532-nm Nd:YAG laser for excitation, and magnetic properties are measured by a superconducting quantum interference device (Quantum Design, SQUID-VSM) magnetometer.

In our study, we subject two types of CrSBr samples to He$^+$ ion irradiation. The first type is a bulk crystal with a thickness of around 200 µm, the top surface of which is irradiated by 1.7-MeV He$^+$ ions under different fluences. We start with an ion fluence of $4\times10^{14}$/cm$^2$ and monitor the ion-induced change in magnetic and structural properties using SQUID magnetometry and Raman spectroscopy. We then systematically repeat this experimental protocol, increasing the ion fluence at each step, until we reach a maximum fluence of $8\times10^{15}$/cm$^2$ (to indicate the fluence, the samples are named as 4E14…8E15). To complement our investigation of this crystal, we study a second type of sample containing many small CrSBr flakes with a thickness below 1 µm that we obtain by exfoliation onto a silicon substrate. To achieve irradiation effects comparable to those in the large bulk crystal, we vary the energy of the He$^+$ ions for the thin flake sample (see SM, Table S1).

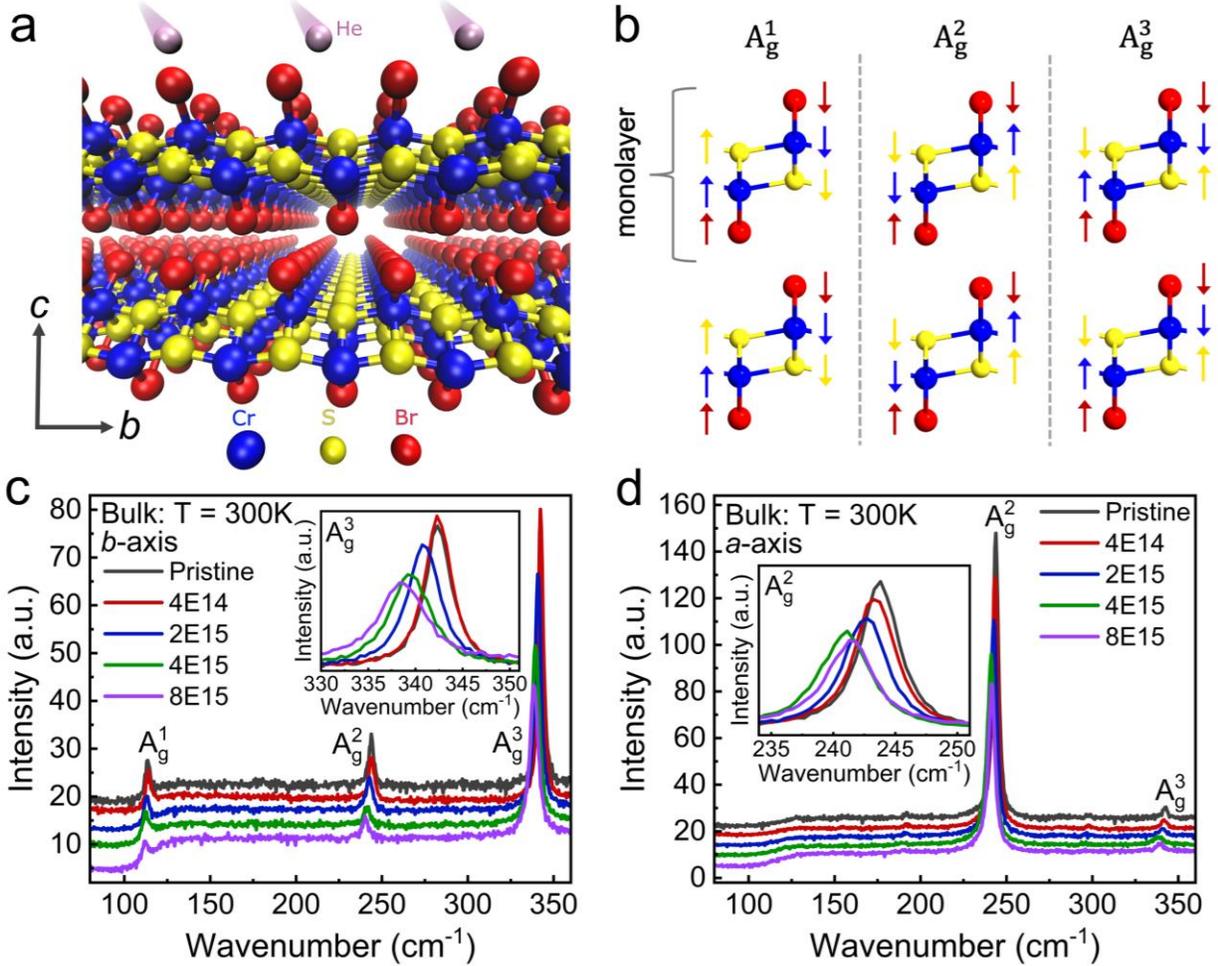

**Figure 1**. Schematic illustration of (a) He-irradiation and (b) lattice vibration modes of CrSBr crystals. Color-coded arrows in (b) indicate the direction along which individual atoms move in the vibration. (c,d) Raman spectra from the 200-µm-thick bulk crystal measured after every ion irradiation step (the numbers correspond to ion fluences, e.g. 4E14 = $4\times10^{14}$/cm$^2$). The laser excitation was polarized along the $b$- and $a$-axis, respectively. The spectra are vertically offset for better visibility. The magnified views in the insets shows a continuous reduction in wavenumber for the $A_g^3$ and $A_g^2$ modes, indicating the gradual softening of the lattice with increasing ion fluence. The measurements were taken at ambient conditions.

As depicted in Fig.1a, each layer of CrSBr is comprised of a buckled plane of CrS complexes surrounded by a sheet of Br atoms. Three characteristic modes, attributed to the $A_g$ modes of the out-of-plane vibration, can be clearly identified in the Raman spectra (cf. Fig. 1b,c,d).[23] The $A_g^1$ (113.8 cm$^{-1}$) and $A_g^3$ (342.5 cm$^{-1}$) modes show maximum intensity when the laser polarization is parallel to the $b$-axis, while the $A_g^2$ (245.2 cm$^{-1}$) mode is most pronounced when the polarization is aligned with the $a$-axis, reflecting the strong structural anisotropy of the crystal structure. Even

CrSBr crystals exposed to the largest ion fluences applied in our study maintain their atomic structure and this anisotropy. The gradual decrease in the peak intensity of the Raman signatures and the continuous shift towards smaller wavenumbers observed in Fig. 1c,d (also see Fig. S1) are generally attributed to a softening of the phonon modes resulting from defect-induced variations in the lattice spacing.[24] This observation is in line with a recent study of defects in mono- and few-layer CrSBr flakes irradiated by He ions, which also demonstrates the structural and electronic anisotropy.[25]

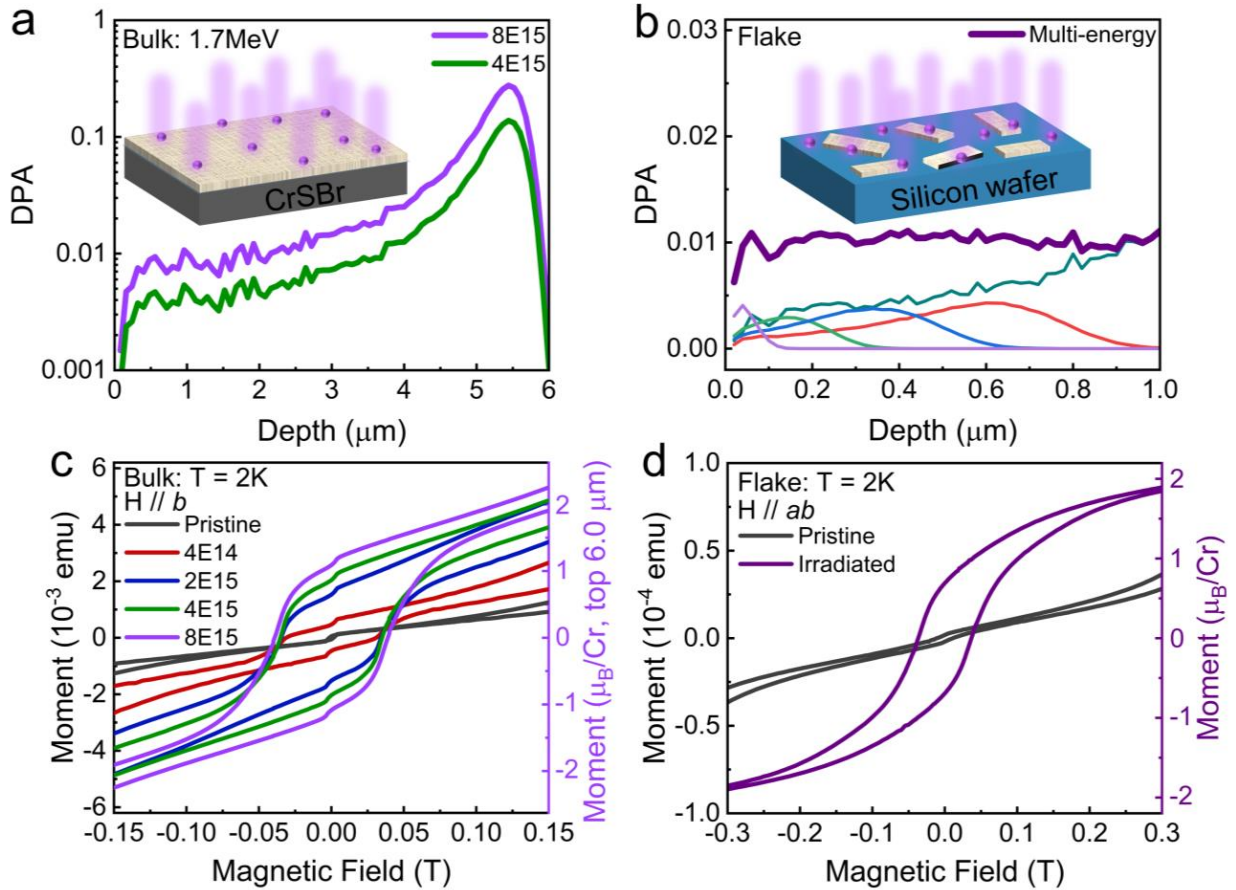

**Figure 2.** DPA for (a) the 200-µm-thick bulk sample with a fluence of $4\times10^{15}/cm^2$ (4E15) and $8\times10^{15}/cm^2$ (8E15); and (b) the flake sample (the thinner lines of different color represent the DPA induced by ions with different energy, see Table S1). The insets schematically show the sample geometry and effective thickness (the beige color). Magnetic hysteresis at 2 K for (c) the 200-µm-thick sample after irradiation at different fluences (the abrupt change near zero field is the antiferromagnetic signal from the un-affected bulk CrSBr) and (d) the flake sample after irradiation with a fluence equivalent to the maximum fluence used for the sample in (c). Measurements taken before irradiation shown for comparison are labeled pristine.

From SRIM (Stopping and Range of Ions in Matter) simulations,[26] we calculate the average number of times an atom is displaced from its equilibrium lattice position during irradiation (displacement per atom, DPA), which allows us to estimate the depth profile of energy transferred from the energetic ions to the crystal matrix and thus the distribution of defects created in the crystal. The DPA of the 200-µm-thick crystal shown in Fig. 2a indicates that a low concentration of defects with relatively homogeneous distribution is created in the top 4 µm of the crystal and that a high concentration of defects is expected to occur 5 to 6 µm below the surface. In the thin flake sample, on the other hand, defects are homogeneously distributed (see Fig. 2b). Overall, we expect a defect concentration comparable to that of the 200-µm-thick crystal created in the top 4 µm.

To investigate the influence of $He^+$ irradiation on the magnetic properties, we measure the field- and temperature-dependent magnetization in the 200-µm-thick bulk crystal and in the thin-flake sample using SQUID-magnetometry. Figure 2c and Figure S2 show the magnetization induced in the bulk crystal by a field applied along the *b*-axis. Before irradiation, the sample exhibits the typical AFM response expected for pristine CrSBr crystals.[7] However, after we expose the crystal to ion irradiation, magnetic hysteresis centered at zero field is observed at fields up to 0.1 T. Upon increasing the irradiation fluence, this fingerprint of FM order becomes more and more prominent. At the same time, at fields around 0.5 T, we also observe a spin-flip transition that is characteristic of AFM-ordered CrSBr (see Fig. S2). This observation is fully in line with the fact that only a fraction of the volume of the 200-µm-thick crystal is altered by ion irradiation. Hence, we observe a signal that has contributions from both irradiated and pristine regions of the sample. For a fluence of $8\times10^{15}/cm^2$, we can estimate that the saturation magnetic moment in the irradiated volume amounts to 2 $\mu_B$/Cr by assuming an irradiation thickness of 6 µm according to Fig. 2a. After 60 days storage at ambient conditions, we re-measured our sample and found no significant change in the magnetic response (see Fig. S3).

The effect of ion irradiation on the magnetic properties is much more prominent in the second sample. For the thin flakes, we observe a full transition of the magnetic ground state from AFM to FM after ion irradiation. As demonstrated by Fig. 2d and the full-field measurements in Fig. S2, magnetic hysteresis centered at zero field dominates the magnetic response after irradiation and the signatures of the AFM spin-flip transition are no longer observed. The saturation magnetization approaches 2 $\mu_B$/Cr. Note that the saturation magnetization of Cr atoms in CrSBr is 3 $\mu_B$/Cr.[27, 28] For the calculation, we assumed that the saturation magnetic moment is 3 $\mu_B$/Cr in the pristine sample and there is no flake loss during sample handling. The in-plane magnetic anisotropy is not observed as a sharp spin-flip transition because individual flakes are randomly oriented on this sample.

In the next step, we investigate irradiation-induced changes of magnetic properties by temperature-dependent magnetization measurements conducted with an applied magnetic field of 0.1 T. Before irradiation, both samples show a peak of the magnetization at the Néel temperature of around 131 K, reflecting the AFM ground state of pristine CrSBr crystals. After irradiation with He$^+$ ions, however, we observe a continuous increases of the magnetization of thin flakes as the temperature decreases (see Fig. 3a), perfectly in line with our expectation of FM order in the irradiated samples. The critical temperature of the FM state is found to be 111 K. The magnetization of the 200-µm-thick bulk sample shows FM contributions as an increase in the absolute magnetization measured at low temperatures, since only a fraction of the volume is affected by ion irradiation. Nonetheless, the difference in the magnetic response after ion irradiation below 120 K is clear in Fig. 3b.

To characterize the remanence of the irradiation-induced FM state, CrSBr samples are cooled down from room temperature while a field of 0.1 T is applied to saturate the FM magnetization. When the temperature reaches 2 K, the magnetic field is set to zero. We then measure the remanence while increasing the temperature. The irradiated 200-µm-thick crystal exhibits sizeable spontaneous magnetization below the critical temperature, as can be clearly seen in Fig. 3(c). We note that, while the magnetization of our bulk sample increases with irradiation fluence, the critical temperature always remains around 110 K for all fluences (see also Fig. S4). Klein et al. suggested a potential defect-related phase after annealing with an onset arising also at around 110 K.[29]

As outlined above, pristine CrSBr shows a magnetic easy axis along the crystallographic *b*-axis, while the *a*- and *c*-axis correspond to intermediate and hard magnetic axes. In another ion-irradiated bulk crystal sample, we measure the temperature-dependent remanence for fields applied along the *a*-, *b*-, and *c*-axis during cool-down. As Fig. S5 shows, the amplitude of the remanence is largest when the field is applied along the *b*-axis. Ion irradiation has changed the magnetic ground state, but the magnetic easy axis is preserved. This result is in good agreement with the Raman measurements presented above, indicating that the crystalline anisotropy of CrSBr remains present even after irradiation with the largest ion fluences used in our experiments.

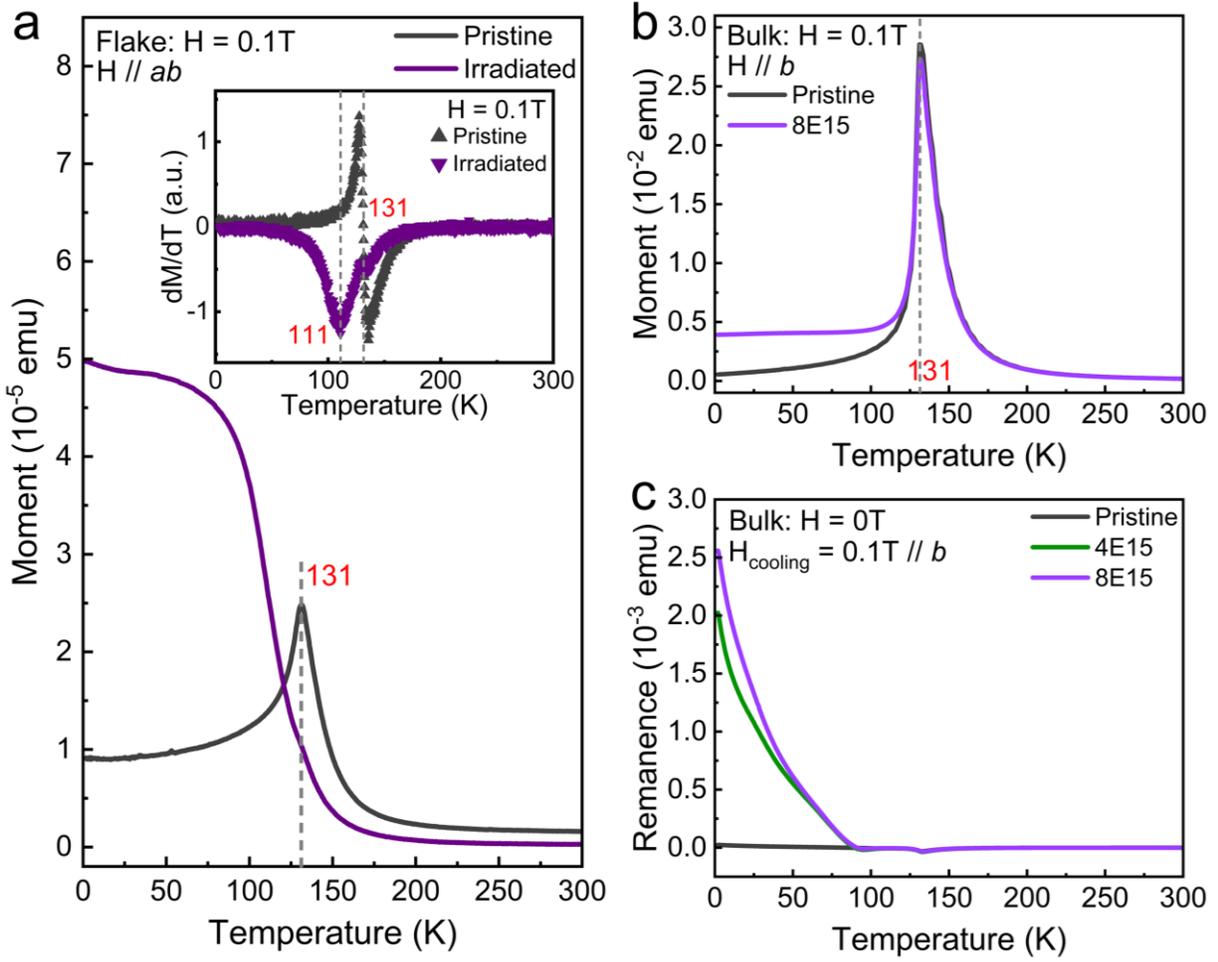

**Figure 3.** Temperature-dependent magnetization of the flake sample (a) and the bulk sample 8E15 along the *b*-axis (b) under an applied magnetic field of 0.1 T. In the inset to (a), the small kink in dM/dT at around 131 K is due to the antiferromagnetic signal of much thicker flakes which are not fully irradiated through their thickness. Measurements on the pristine crystals are also shown for comparison. (c) Magnetic remanence of the 200-µm-thick bulk sample for two different fluences.

The transfer of energy from high-energy ions to atoms in the crystal lattice leads to the formation of defects with very high concentrations. In our case here, the SRIM simulation results in a defect concentration of around 1% for the flake sample and the top 4 µm of the bulk sample (see Figs. 2a, b). The most common type of defects is assumed to be Frenkel pairs – a vacancy and an interstitial atom – which are created when an atom is displaced from its equilibrium position in the lattice into an interstitial position, leaving behind an empty lattice site. This kind of lattice disorder has been shown to modify conventional 3D magnetic materials.[30, 31]

We calculated the energetics of different types of vacancies and interstitials using 4×4 CrSBr bilayer supercells corresponding to a defect concentration of ~0.5%. The interstitial atoms were initially placed at different positions and the most energetically favorable structure was identified. The formation energies for Cr, S, and Br vacancies and interstitials are summarized in Table 1. Assuming that vacancies and interstitials are not interacting with each other, the Frenkel pair formation energy can be found by adding the formation energies of the isolated defects. Although the Frenkel pair formation energy is similar for three different elements, the lower values were found for S and Br, suggesting a higher concentration of these types of defects in the material. Br vacancies in CrSBr have been identified via scanning tunneling microscopy in a recent work,[29] which is consistent with our lowest vacancy formation energy found for Br (3.59 eV/atom). Table S2 lists the total magnetic moment for various vacancies and interstitials. While the net magnetic moment in pristine CrSBr is zero, our calculations show a total magnetic moment of 2.0, 6.04, and 1.05 $\mu_B$/supercell for S, Cr, and Br vacancies, respectively. In contrast, the S interstitial does not significantly change the magnetic moments of the pristine bilayer. The presence of magnetic character can be related to the shift between the spin-up- and spin-down-associated states in the electronic structure of the defective materials, as presented in Figs. S6 and S7.

We further investigated the preferred magnetic ordering including the energy difference between the lowest energy AFM and FM ordering (see Fig. 4). For all types of interstitials, the moments of the interstitial atoms are aligned with the Cr atoms at the bottom and top, resulting in the energetically favorable FM order. Although the energy difference depends on the exchange parameter values, the FM preference was still found for various exchange values (Table S3). The effect is more pronounced in CrSBr bilayer with Cr interstitials. The optimized structure of the CrSBr bilayer with Cr interstitial (Fig. S8) indicates that the interstitial atom chooses a position in the void between the van der Waals gap atoms forming a covalent bond between the layers, which is consistent with the previous results.[32] The first-nearest-neighbor exchange interactions between Cr-Br-Cr and Cr-S-Cr, as well as the second-nearest-neighbor exchange between Cr-S-Cr interactions, are the main causes of the ferromagnetism of the CrSBr monolayer.[33] The former exchange interaction is mediated by the orbital hybridization between Cr interstitials and inner Br atoms (Fig. S9). As a result, the AFM coupling strength between the layers becomes weaker because of the appearance of defects.

|   | $E_V$ (eV) | $E_I$ (eV) | $E_F$ (eV) |
|---|---|---|---|
| S | 5.89 | -2.18 | 3.71 |
| Cr | 7.76 | -2.81 | 4.95 |
| Br | 3.59 | 0.45 | 4.03 |

**Table 1.** Results of a calculation of the formation energies of different vacancies ($E_V$), interstitials ($E_I$), and Frenkel ($E_F$) defects in CrSBr bilayer structures.

We note that the change of the magnetic ground state we observe after ion irradiation could in principle also be caused by lattice expansion, e.g., due to intercalation[9]. However, X-ray diffraction and cross-sectional high-resolution scanning transmission electron microscopy measurements of our irradiated crystals (see Figs. S10, S11) do not indicate any expansion of the lattice along the $c$-axis within the detection limit. Recent works have demonstrated the possibility of strain, hydrostatic pressure or structural phase transition in modifying the magnetic properties of CrSBr.[19, 22, 32]

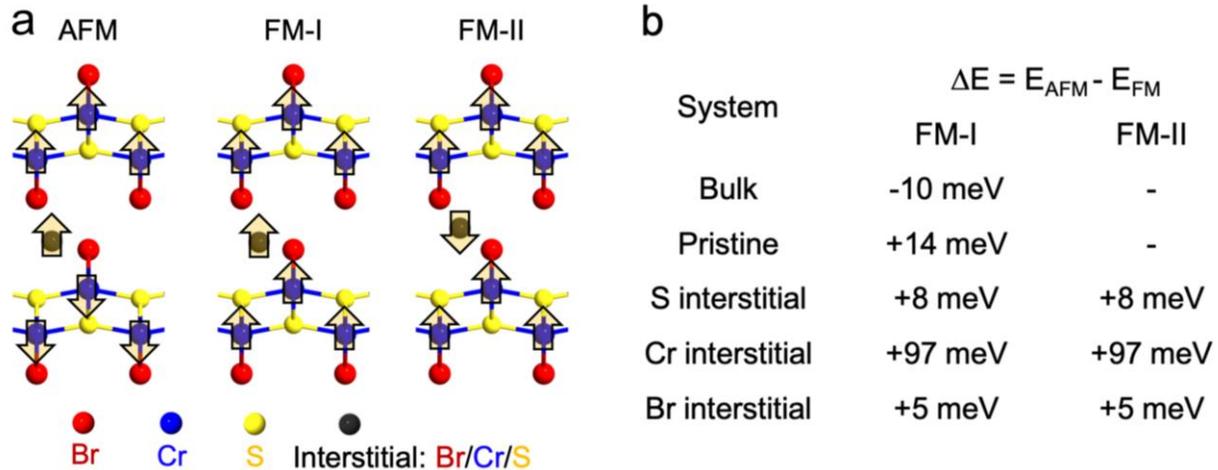

**Figure 4**. (a) Schematic representation of the magnetic configurations of interstitials in the CrSBr crystal. Regular Cr, S, and Br atoms are represented by smaller blue, yellow, and red circles, respectively. Orientations of the intralayer Cr spins are depicted by the arrows. Three spin configurations are considered for the bilayer CrSBr with interstitials. In the FM case, the magnetic moments of the interstitials can be aligned (FM-I) or anti-aligned (FM-II) with respect to those in the bottom and top layer; (b) The energy difference $\Delta E = E_{AFM} - E_{FM}$ between the AFM and FM configuration for the bulk system and bilayer with/without interstitials is given for the supercell. Interstitial density corresponds to ~0.5%. A positive $\Delta E$ indicates that FM is favored over AFM configuration.

We observe a change of the magnetic ground state of the vdW magnet CrSBr after irradiation by non-magnetic ions. The fluence-dependent remanent magnetization of the irradiation-induced

FM state exists up to a critical temperature of 110 K and adapts the magnetic easy axis anisotropy of the AFM order before irradiation. The saturation magnetic moment of the FM magnetization is estimated to be close to the maximum magnetic moment of the constituent Cr atoms. First-principles calculations suggest that the displacement of atoms into interstitial positions by high-energy ions is responsible for the experimentally observed transition of the magnetic ground state, as the interstitials facilitate ferromagnetic coupling between the layers (structural units) of CrSBr. We note that ion implantation is a mature technology for chip fabrication. Applying proper lithography, one can create artificial lateral structures. By tuning the energy of the ions and hence the penetration depth, it may further be possible, in a vdW heterostructure, to maximize the impact of ions for CrSBr but not the material on top. As a result, our approach may be applied to modify the magnetic properties of CrSBr even after the fabrication of a vdW heterostructure.

## Supporting Information

The Supporting Information is available free of charge at xxxx.

Methods, He irradiation parameters for the flake sample, Calculated magnetic moments for various vacancies and interstitials, The energy difference between the AFM and FM configurations, Evolution of the Raman modes of $A_g^1$ and $A_g^2$ as a function of irradiation fluence, Magnetization curves at 2 K with a large field range, Magnetization curves at 2K for irradiated flake sample and after being exposed to the ambient condition for 60 days, Magnetic remanence of the bulk sample with different irradiation fluences, Magnetic remanence of the bulk sample 8E15 along different crystallographic directions, Projected density of states (PDOS) of pristine and defective CrSBr monolayers with S Cr, and Br vacancies, Projected density of states (PDOS) of pristine and defective CrSBr monolayers with S Cr, and Br interstitials, Optimized structures of CrSBr bilayer with different interstitials, Projected density of states for CrSBr bilayer with Cr interstitial in FM and AFM ordering, X-ray diffraction of pristine CrSBr and sample 8E15, Cross-sectional high-resolution HAADF-STEM images and corresponding fast Fourier transforms.

## Acknowledgement

F.D. gratefully acknowledges financial support from Alexey Chernikov and the Würzburg-Dresden Cluster of Excellence on Complexity and Topology in Quantum Matter ct.qmat (EXC 2147, Project-ID 390858490). A.V.K. thanks the German Research Foundation (DFG) through projects KR 4866/9-1 and the collaborative research center "Chemistry of Synthetic 2D Materials" CRC-

1415-417590517. Generous CPU time grants from the Technical University of Dresden computing cluster (TAURUS) and Gauss Centre for Supercomputing e.V. (www.gauss-centre.eu), Supercomputer HAWK at Höchstleistungsrechenzentrum Stuttgart (www.hlrs.de), are greatly appreciated. A. K. acknowledges financial support from the Spanish Ministry for Science and Innovation – AEI Grant CEX2018-000805-M (through the "Maria de Maeztu" Programme for Units of Excellence in R&D) and grant RYC2021-031063-I funded by MCIN/AEI/10.13039/501100011033 and "European Union Next Generation EU/PRTR". The authors thank Annette Kunz for TEM specimen preparation. Ion irradiation was performed at the Ion Beam Center (IBC) of HZDR. The use of the IBC TEM facilities and the funding of TEM Talos by the German Federal Ministry of Education and Research (BMBF), Grant No. 03SF0451, in the framework of HEMCP are acknowledged. F.L. thanks the financial support from China Scholarship Council (File No. 202108440218) for his stay in Germany. Z.S. was supported by ERC-CZ program (project LL2101) from Ministry of Education Youth and Sports (MEYS) and used large infrastructure from project reg. No. CZ.02.1.01/0.0/0.0/15_003/0000444 financed by the EFRR).

References

(1) Sierra, J. F.; Fabian, J.; Kawakami, R. K.; Roche, S.; Valenzuela, S. O. Van der Waals heterostructures for spintronics and opto-spintronics. *Nat. Nanotechnol.* **2021,** 16 (8), 856-868.
(2) Huang, B.; Clark, G.; Navarro-Moratalla, E.; Klein, D. R.; Cheng, R.; Seyler, K. L.; Zhong, D.; Schmidgall, E.; McGuire, M. A.; Cobden, D. H.; Yao, W.; Xiao, D.; Jarillo-Herrero, P.; Xu, X. Layer-dependent ferromagnetism in a van der Waals crystal down to the monolayer limit. *Nature* **2017,** 546 (7657), 270-273.
(3) Gong, C.; Li, L.; Li, Z.; Ji, H.; Stern, A.; Xia, Y.; Cao, T.; Bao, W.; Wang, C.; Wang, Y.; Qiu, Z. Q.; Cava, R. J.; Louie, S. G.; Xia, J.; Zhang, X. Discovery of intrinsic ferromagnetism in two-dimensional van der Waals crystals. *Nature* **2017,** 546 (7657), 265-269.
(4) Rahman, S.; Liu, B.; Wang, B.; Tang, Y.; Lu, Y. Giant Photoluminescence Enhancement and Resonant Charge Transfer in Atomically Thin Two-Dimensional $Cr_2Ge_2Te_6$/$WS_2$ Heterostructures. *ACS Appl. Mater. Interfaces* **2021,** 13 (6), 7423-7433.
(5) Shcherbakov, D.; Stepanov, P.; Weber, D.; Wang, Y.; Hu, J.; Zhu, Y.; Watanabe, K.; Taniguchi, T.; Mao, Z.; Windl, W.; Goldberger, J.; Bockrath, M.; Lau, C. N. Raman Spectroscopy, Photocatalytic Degradation, and Stabilization of Atomically Thin Chromium Tri-iodide. *Nano Lett.* **2018,** 18 (7), 4214-4219.
(6) Long, G.; Henck, H.; Gibertini, M.; Dumcenco, D.; Wang, Z.; Taniguchi, T.; Watanabe, K.; Giannini, E.; Morpurgo, A. F. Persistence of Magnetism in Atomically Thin $MnPS_3$ Crystals. *Nano Lett.* **2020,** 20 (4), 2452-2459.
(7) Telford, E. J.; Dismukes, A. H.; Dudley, R. L.; Wiscons, R. A.; Lee, K.; Chica, D. G.; Ziebel, M. E.; Han, M. G.; Yu, J.; Shabani, S.; Scheie, A.; Watanabe, K.; Taniguchi, T.; Xiao, D.; Zhu, Y.; Pasupathy, A. N.; Nuckolls, C.; Zhu, X.; Dean, C. R.; Roy, X. Coupling between magnetic order and charge transport in a two-dimensional magnetic semiconductor. *Nat. Mater.* **2022,** 21 (7), 754-760.

(8) Göser, O.; Paul, W.; Kahle, H.G. Magnetic properties of CrSBr. *J. Magn. Magn. Mater.* **1990,** 92, 129-136.
(9) Pereira, J. M.; Tezze, D.; Ormaza, M.; Hueso, L. E.; Gobbi, M. Engineering Magnetism and Superconductivity in van der Waals Materials via Organic-Ion Intercalation. *Advanced Physics Research* **2023**, 2, 2200084.
(10) Wang, Q. H.; Bedoya-Pinto, A.; Blei, M.; Dismukes, A. H.; Hamo, A.; Jenkins, S.; Koperski, M.; Liu, Y.; Sun, Q. C.; Telford, E. J.; Kim, H. H.; Augustin, M.; Vool, U.; Yin, J. X.; Li, L. H.; Falin, A.; Dean, C. R.; Casanova, F.; Evans, R. F. L.; Chshiev, M.; Mishchenko, A.; Petrovic, C.; He, R.; Zhao, L.; Tsen, A. W.; Gerardot, B. D.; Brotons-Gisbert, M.; Guguchia, Z.; Roy, X.; Tongay, S.; Wang, Z.; Hasan, M. Z.; Wrachtrup, J.; Yacoby, A.; Fert, A.; Parkin, S.; Novoselov, K. S.; Dai, P.; Balicas, L.; Santos, E. J. G. The Magnetic Genome of Two-Dimensional van der Waals Materials. *ACS Nano* **2022,** 16 (5), 6960-7079.
(11) Wilson, N. P.; Lee, K.; Cenker, J.; Xie, K.; Dismukes, A. H.; Telford, E. J.; Fonseca, J.; Sivakumar, S.; Dean, C.; Cao, T.; Roy, X.; Xu, X.; Zhu, X. Interlayer electronic coupling on demand in a 2D magnetic semiconductor. *Nat. Mater.* **2021,** 20 (12), 1657-1662.
(12) Klein, J.; Pingault, B.; Florian, M.; Heissenbuttel, M. C.; Steinhoff, A.; Song, Z.; Torres, K.; Dirnberger, F.; Curtis, J. B.; Weile, M.; Penn, A.; Deilmann, T.; Dana, R.; Bushati, R.; Quan, J.; Luxa, J.; Sofer, Z.; Alu, A.; Menon, V. M.; Wurstbauer, U.; Rohlfing, M.; Narang, P.; Loncar, M.; Ross, F. M. The Bulk van der Waals Layered Magnet CrSBr is a Quasi-1D Material. *ACS Nano* **2023**, 17 (6), 5316-5328.
(13) Ye, C.; Wang, C.; Wu, Q.; Liu, S.; Zhou, J.; Wang, G.; Soll, A.; Sofer, Z.; Yue, M.; Liu, X.; Tian, M.; Xiong, Q.; Ji, W.; Renshaw Wang, X. Layer-Dependent Interlayer Antiferromagnetic Spin Reorientation in Air-Stable Semiconductor CrSBr. *ACS Nano* **2022,** 16, 11876-11883.
(14) Dirnberger, F.; Quan, J.; Bushati, R.; Diederich, G. M.; Florian, M.; Klein, J.; Mosina, K.; Sofer, Z.; Xu, X.; Kamra, A.; García-Vidal, F. J.; Alù, A.; Menon, V. M. Magneto-optics in a van der Waals magnet tuned by self-hybridized polaritons. *Nature* **2023**, 620, 533–537.
(15) Telford, E. J.; Dismukes, A. H.; Lee, K.; Cheng, M.; Wieteska, A.; Bartholomew, A. K.; Chen, Y. S.; Xu, X.; Pasupathy, A. N.; Zhu, X.; Dean, C. R.; Roy, X. Layered Antiferromagnetism Induces Large Negative Magnetoresistance in the van der Waals Semiconductor CrSBr. *Adv. Mater.* **2020,** 32 (37), 2003240.
(16) Y. Guo; Y. Zhang; S. Yuan; B. Wang; J. Wang. Chromium sulfide halide monolayers: intrinsic ferromagnetic semiconductors with large spin polarization and high carrier mobility. *Nanoscale* **2018,** 10 (37), 18036-18042.
(17) Lee, K.; Dismukes, A. H.; Telford, E. J.; Wiscons, R. A.; Wang, J.; Xu, X.; Nuckolls, C.; Dean, C. R.; Roy, X.; Zhu, X. Magnetic Order and Symmetry in the 2D Semiconductor CrSBr. *Nano Lett.* **2021,** 21 (8), 3511-3517.
(18) Boix-Constant, C.; Manas-Valero, S.; Ruiz, A. M.; Rybakov, A.; Konieczny, K. A.; Pillet, S.; Baldovi, J. J.; Coronado, E. Probing the Spin Dimensionality in Single-Layer CrSBr Van Der Waals Heterostructures by Magneto-Transport Measurements. *Adv. Mater.* **2022,** 34 (41), 2204940.
(19) Cenker, J.; Sivakumar, S.; Xie, K.; Miller, A.; Thijssen, P.; Liu, Z.; Dismukes, A.; Fonseca, J.; Anderson, E.; Zhu, X.; Roy, X.; Xiao, D.; Chu, J. H.; Cao, T.; Xu, X. Reversible strain-induced magnetic phase transition in a van der Waals magnet. *Nat. Nanotechnol.* **2022,** 17 (3), 256-261.
(20) Cenker, J.; Ovchinnikov, D.; Yang, H.; Chica, D. G.; Zhu, C.; Cai, J. Q.; Diederich, G.; Liu, Z. Y.; Zhu, X. Y.; Roy, X.; Cao, T.; W., M.; Daniels; Chu, J.-H.; Xiao, D.; Xu, X. D. Strain-programmable van der Waals magnetic tunnel junctions. *arXiv (Condensed Matter)*, Janurary 10, 2023, 2301.03759. DOI: 10.48550/arXiv.2301.03759.
(21) Telford, E. J.; Chica, D. G.; Xie, K. C.; Manganaro, N. S.; Huang, C.-H.; Cox, J.; Dismukes, A. H.; Zhu, X. Y.; Walsh, J. P. S.; Cao, T.; Dean, C. R.; Roy, X.; Ziebel, M. E. Designing magnetic properties in CrSBr through hydrostatic pressure and ligand substitution. *arXiv (Condensed Matter)*, November 5, 2022, 2211.02788. DOI: 10.48550/arXiv.2211.02788.
(22) Pawbake, A.; Pelini, T.; Mohelsky, I.; Jana, D.; Breslavetz, I.; Cho, C.-W.; Orlita, M.; Potemski, M.; Measson, M.-A.; Wilson, N.; Mosina, K.; Soll, A.; Sofer, Z.; Piot, B. A.; Zhitomirsky, M. E.; Faugeras, C. Magneto-optical sensing of the pressure driven magnetic ground states in bulk CrSBr. *arXiv (Condensed Matter)*, March 3, 2023, 2303.01823. DOI: 10.48550/arXiv.2303.01823.

(23) Moro, F.; Ke, S.; del Águila, A. G.; Söll, A.; Sofer, Z.; Wu, Q.; Yue, M.; Li, L.; Liu, X.; Fanciulli, M. Revealing 2D Magnetism in a Bulk CrSBr Single Crystal by Electron Spin Resonance. *Adv. Funct. Mater.* **2022,** 32, 2207044.

(24) Parkin, W. M.; Balan, A.; Liang, L.; Das, P. M.; Lamparski, M.; Naylor, C. H.; Rodriguez-Manzo, J. A.; Johnson, A. T.; Meunier, V.; Drndic, M. Raman Shifts in Electron-Irradiated Monolayer $MoS_2$. *ACS Nano* **2016,** 10 (4), 4134-4142.

(25) Torres, K.; Kuc, A.; Maschio, L.; Pham, T.; Reidy, K.; Dekanovsky, L.; Sofer, Z.; Ross, F. M.; Klein, J. Probing Defects and Spin-Phonon Coupling in CrSBr via Resonant Raman Scattering. *Adv. Funct. Mater.* **2023,** 33 (12), 2211366.

(26) Ziegler, J. F.; Ziegler, M. D.; Biersack, J. P. SRIM – The stopping and range of ions in matter (2010). *Nucl. Instr. and Meth. B* **2010,** 268 (11-12), 1818-1823.

(27) Jiang, Z.; Wang, P.; Xing, J.; Jiang, X.; Zhao, J. Screening and Design of Novel 2D Ferromagnetic Materials with High Curie Temperature above Room Temperature. *ACS Appl. Mater. Interfaces* **2018,** 10 (45), 39032-39039.

(28) Lopez-Paz, S. A.; Guguchia, Z.; Pomjakushin, V. Y.; Witteveen, C.; Cervellino, A.; Luetkens, H.; Casati, N.; Morpurgo, A. F.; von Rohr, F. O. Dynamic magnetic crossover at the origin of the hidden-order in van der Waals antiferromagnet CrSBr. *Nat. Commun.* **2022,** 13 (1), 4745-4754.

(29) J. Klein; Z. Song; B. Pingault; F. Dirnberger; H. Chi; J. B. Curtis; R. Dana; R. Bushati; J. Quan; L. Dekanovsky; Z. Sofer; A. Alù; V. M. Menon; J. S. Moodera; M. Lončar; P. Narang; F. M. Ross. Sensing the local magnetic environment through optically active defects in a layered magnetic semiconductor. *ACS Nano* **2023,** 17, 288-299.

(30) Fassbender, J. Nanopatterning: the chemical way to ion irradiation. *Nat. Nanotechnol.* **2012,** 7 (9), 554-555.

(31) Nord, M.; Semisalova, A.; Kakay, A.; Hlawacek, G.; MacLaren, I.; Liersch, V.; Volkov, O. M.; Makarov, D.; Paterson, G. W.; Potzger, K.; Lindner, J.; Fassbender, J.; McGrouther, D.; Bali, R. Strain Anisotropy and Magnetic Domains in Embedded Nanomagnets. *Small* **2019,** 15 (52), 1904738.

(32) Klein, J.; Pham, T.; Thomsen, J. D.; Curtis, J. B.; Denneulin, T.; Lorke, M.; Florian, M.; Steinhoff, A.; Wiscons, R. A.; Luxa, J.; Sofer, Z.; Jahnke, F.; Narang, P.; Ross, F. M. Control of structure and spin texture in the van der Waals layered magnet CrSBr. *Nat. Commun.* **2022,** 13 (1), 5420-5428.

(33) Xu, X. M.; Wang, X. H.; Chang, P.; Chen, X. Y.; Guan, L. X.; Tao, J. G. Strong Spin-Phonon Coupling in Two-Dimensional Magnetic Semiconductor CrSBr. *J. Phys. Chem. C* **2022,** 126 (25), 10574-10583.

TOC:

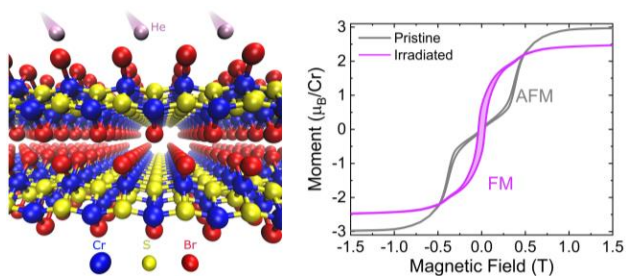

# Ferromagnetic interlayer coupling in CrSBr crystals irradiated by ions

Fangchao Long[1,2], Mahdi Ghorbani-Asl[1], Kseniia Mosina[3], Yi Li[1,2], Kaiman Lin[1,4], Fabian Ganss[1], René Hübner[1], Zdenek Sofer[3], Florian Dirnberger[5*], Akashdeep Kamra[6], Arkady V. Krasheninnikov[1], Slawomir Prucnal[1], Manfred Helm[1,2], Shengqiang Zhou[1*]

[1]Helmholtz-Zentrum Dresden-Rossendorf, Institute of Ion Beam Physics and Materials Research, Bautzner Landstrasse 400, 01328 Dresden, Germany

[2]Technische Universität Dresden, 01069 Dresden, Germany

[3]Department of Inorganic Chemistry, University of Chemistry and Technology Prague, Technická 5, 166 28 Prague 6, Czech Republic

[4]University of Michigan-Shanghai Jiao Tong University Joint Institute, Shanghai Jiao Tong University, Shanghai, China

[5]Institute of Applied Physics and Würzburg-Dresden Cluster of Excellence ct.qmat, Technische Universität Dresden, Germany

[6]Condensed Matter Physics Center (IFIMAC) and Departamento de Física Teórica de la Materia Condensada, Universidad Autónoma de Madrid, Madrid, Spain

- Corresponding authors: florian.dirnberger@tu-dresden.de, s.zhou@hzdr.de

# Supporting Information

**Table of Contents**

Figure S5. Magnetic remanence of the bulk sample 8E15 along different crystallographic directions

Figure S6. Projected density of states (PDOS) of pristine and defective CrSBr monolayers with S Cr, and Br vacancies

Figure S7. Projected density of states (PDOS) of pristine and defective CrSBr monolayers with S Cr, and Br interstitials

Figure S8. Optimized structures of CrSBr bilayer with different interstitials

Figure S9. Projected density of states for CrSBr bilayer with Cr interstitial in FM and AFM ordering

Figure S10. X-ray diffraction of pristine CrSBr and sample 8E15

Figure S11. Cross-sectional high-resolution HAADF-STEM images and corresponding fast Fourier transforms

## Methods

*Crystal growth*

CrSBr crystals were prepared by direct reaction from the elements. Chromium (99.99%, -60 mesh, Chemsavers, USA), bromine (99.9999%, Sigma-Aldrich, Czech Republic), and sulfur (99.9999%, Stanford Materials, USA) were mixed in stochiometric ratio in a quartz ampoule (35 x 220 mm) corresponding to 15 g of CrSBr. Bromine excess of 0.5 g was used to enhance vapor transport. The material was pre-reacted in an ampoule using a crucible furnace at 700 °C for 12 hours, while the second end of the ampoule was kept below 250 °C. The heating procedure was repeated two times until the liquid bromine disappeared. Afterwards, the ampoule was placed in a horizontal two-zone furnace for crystal growth. First, the growth zone was heated to 900 °C, while the source zone was heated to 700 °C for 25 hours. For the growth, the thermal gradient was reversed and the source zone was heated from 900 °C to 940 °C and the growth zone from 850 °C to 800°C over a period of 7 days. The crystals with dimensions of up to 5 x 20 mm² were removed from the ampoule in an Ar glovebox.

*Flake exfoliation procedure*

First, we picked up a CrSBr crystal and put it on the scotch tape. Then, we used plastic tweezers to crumble the CrSBr crystal to flakes and distributed them evenly within an area of 4 mm × 4 mm. Finally, we put a clean, naturally oxidized Si wafer on the scotch tape containing the CrSBr flakes and pressed both together. After a few minutes, we removed the Si wafer from the tape and got the prepared sample.

*He irradiation procedure*

The CrSBr bulk sample was pasted on a piece of Si wafer. For ion irradiation, a He beam of 1.7 MeV energy with a beam size of 2 mm in diameter produced by a van de Graff accelerator was used. For the flake samples, the He ion beam with different energies (see Table S1) was scanned across the sample surface.

For the bulk sample, we estimated the irradiation volume according to the SRIM simulation result. The SRIM calculation was done by using the option "Quick Calculation of Damage". A default displacement energy of 25 eV was assumed for all elements, since, according to our knowledge, there is no literature value available for CrSBr. Our simulation is only to give a rough picture of the upper limit of damages and their depth profile. The effective depth here is about 6 μm, and the surface area of CrSBr crystal is about 1 mm × 2 mm. However, for the flake sample we cannot measure the mass and area since it contains numerous different flakes. We assumed that the saturation magnetic moment is 3 $\mu_B$/Cr in the pristine sample to calculate the sample mass. Then, we further assumed no flake loss during sample handling and calculated the saturation magnetization after irradiation from the mass we obtained from the pristine flake sample.

*Computational Methods*

The energetics and magnetic properties of all point defects were investigated using spin-polarized density functional theory (DFT) as implemented in the VASP code.[1,2] All the calculations were carried out using the Perdew-Burke-Ernzerhof exchange-correlation functional.[3] The structural models were fully optimized with an energy cut-off of 500 eV and a force tolerance of 0.01 eV Å$^{-1}$. van der Waals (vdW) interactions were taken into account using the Grimme method (DFT-D3)".[4] The point defects were modeled using a 4 × 4 × 1 supercell with a vacuum space of 20 Å in the confinement direction. The Brillouin zone of the primitive cells was sampled using 8 × 8 × 1 k-points. The electronic structure calculations were performed using the DFT + U method with an effective Hubbard value (U) of 3.0 eV for Cr atoms. The energetics of vacancies and interstitials were assessed as $E_f = E_{defective} - (E_{pristine} \pm n_x\mu_x)$, where $E_{defective}$ and $E_{pristine}$ are the energies of the defective and pristine supercell, respectively. $n_x$ is the number of vacancy or interstitial atoms, and $\mu_x$ represents the chemical potential of the X species, which is considered to be the energy of the isolated atom.

*Characterization Methods*

Structural characterization of the as-synthesized CrSBr was performed by micro-Raman spectroscopy using a linearly polarized continuous 532 nm Nd:YAG laser for excitation. The magnetization data was collected by a superconducting quantum interference device (Quantum Design, SQUID-VSM) magnetometer.

We have carried out additional structural characterization for selected samples by X-ray diffraction of Cu K$_{\alpha1}$ radiation on a Malvern Panalytical Empyrean and cross-sectional high-resolution scanning transmission electron microscopy (STEM) imaging with a high-angle annular dark-field (HAADF) detector employing a Talos F200X microscope (Thermo Fisher) operated at an accelerating voltage of 200 kV.

**Table S1.** The parameters of He irradiation for the flake sample: to achieve a homogenous *dpa* (displacement per atom) in the top around 1 µm, we applied multiple energies with different fluences. The resulted *dpa* is similar as for the bulk sample 8E15.

|   | *Energy (keV)* | *Fluence (cm$^{-2}$)* |
|---|---|---|
| *1* | 390 | $1\times10^{15}$/cm$^2$ |
| *2* | 100 | $1\times10^{14}$/cm$^2$ |
| *3* | 50 | $8\times10^{13}$/cm$^2$ |
| *4* | 20 | $5\times10^{13}$/cm$^2$ |
| *5* | 5 | $5\times10^{13}$/cm$^2$ |

**Table S2.** Calculated magnetic moments for 3 different vacancies and interstitials in CrSBr bilayer AFM structures. The total magnetic moments are given per vacancy/interstitial for the supercell.

|   | *S* | | *Cr* | | *Br* | |
|---|---|---|---|---|---|---|
|   | Vacancy | Interstitial | Vacancy | Interstitial | Vacancy | Interstitial |
| *Magnetic moment ($\mu_B$)* | 2.0 | 0.0 | 6.04 | 3.17 | 1.05 | 0.99 |

**Table S3.** Energy difference $\Delta E = E_{AFM} - E_{FM}$ between the AFM and FM configuration is given for different on-site exchange ($J$) parameters of DFT+U calculations. The interstitial density is 0.5%. Here a positive sign indicates an energetically more favorable ferromagnetic ordering.

| Type of system | Hubbard parameters | $\Delta E = E_{AFM} - E_{FM}$ |
|---|---|---|
| Bulk (pristine) | J = 0, U = 3 | -10 meV |
|  | J = 1, U = 3 | 2 meV |
|  | J = 2, U = 3 | 2 meV |
| Bilayer (pristine) | J = 0, U = 3 | 14 meV |
|  | J = 1, U = 3 | 10 meV |
|  | J = 2, U = 3 | 7 meV |
| Bilayer with Cr-interstitial | J = 0, U = 3 | 97 meV |
|  | J = 1, U = 3 | 97 meV |
|  | J = 2, U = 3 | 47 meV |

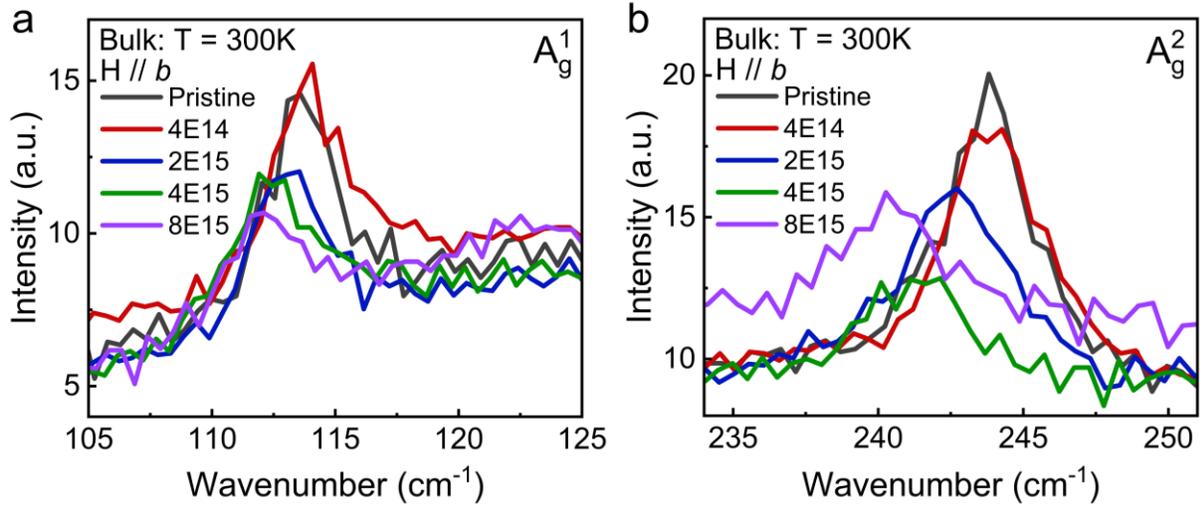

**Figure S1.** Evolution of the Raman modes $A_g^1$ and $A_g^2$ as a function of irradiation fluence.

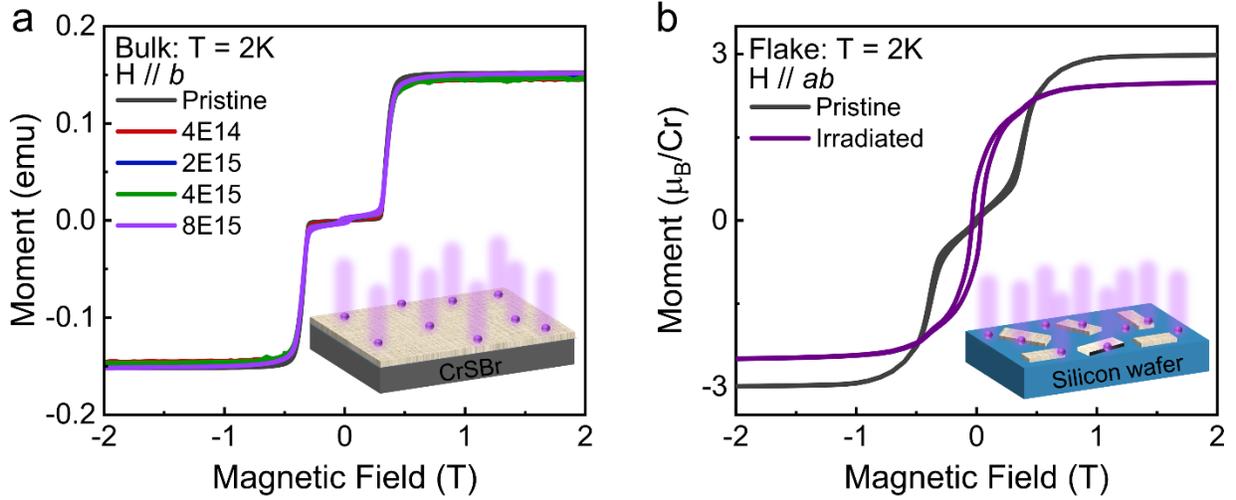

**Figure S2.** Magnetization curves (MH) at 2 K with a large field range for (a) the bulk sample before and after irradiation at different fluences; and (b) the flake sample before and after irradiation at an equivalent fluence of $8 \times 10^{15}/cm^2$.

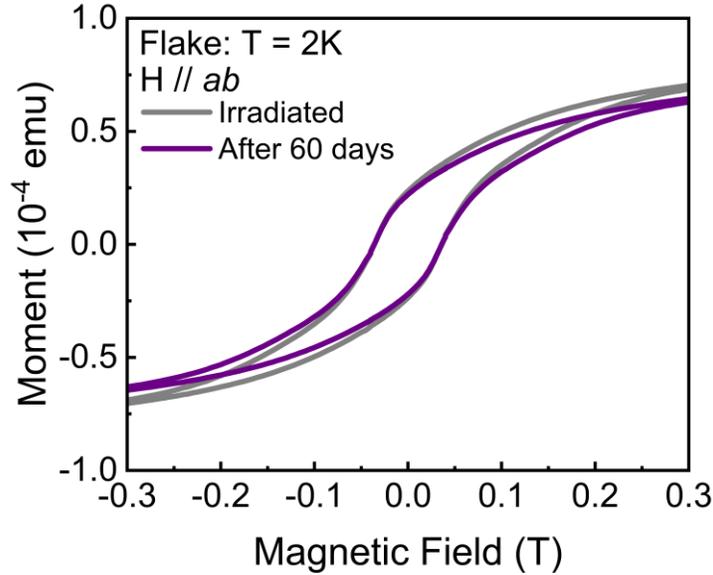

**Figure S3.** Magnetization curves (MH) at 2 K for the irradiated flake sample and after being exposed to the ambient condition for 60 days. There is no significant decay.

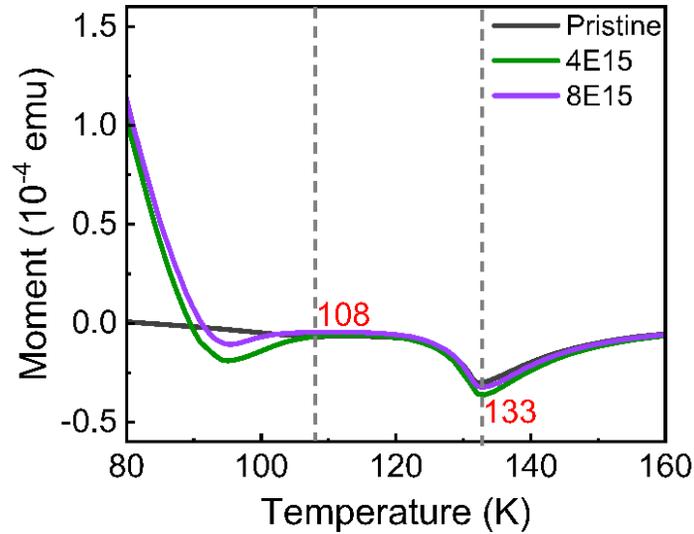

**Figure S4.** Magnetic remanence (the zoom-in near the critical temperature) of the bulk sample with different irradiation fluences along the *b*-axis. For the measurements, the sample was cooled down from room temperature to 2 K under the magnetic field 0.1 T along the *b*-axis, then the field was set to zero, and the magnetic remanence was measured during warming up. In the temperature range 90-110 K, one can see a weak dip, which is due to the possible negative residual magnetic field in the superconducting magnet (when being decreased from +1000 Oe to "0"), probably a few Oe. Near the critical temperature, this residual field is enough to flip some ferromagnetic nano-regions. Due to the same origin, the pronounced dip at 133 K corresponds to the Néel temperature of the non-irradiated bulk crystal. For both, bulk and flake samples, we estimated the critical temperature of the induced ferromagnetism to around 110 K.

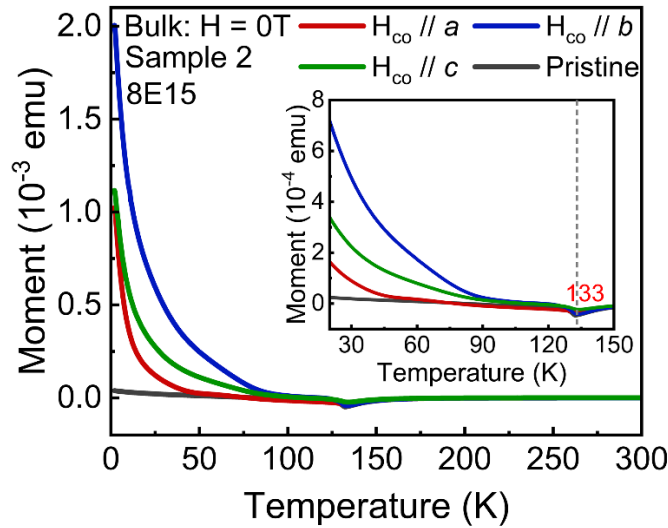

**Figure S5.** Magnetic remanence of the bulk sample 8E15 (a second sample) along different crystallographic directions. For the measurements, the sample was cooled down from room temperature to 2 K under the magnetic field $H_{co}$=0.1 T, then the field was set to zero, and the magnetic remanence was measured during warming up.

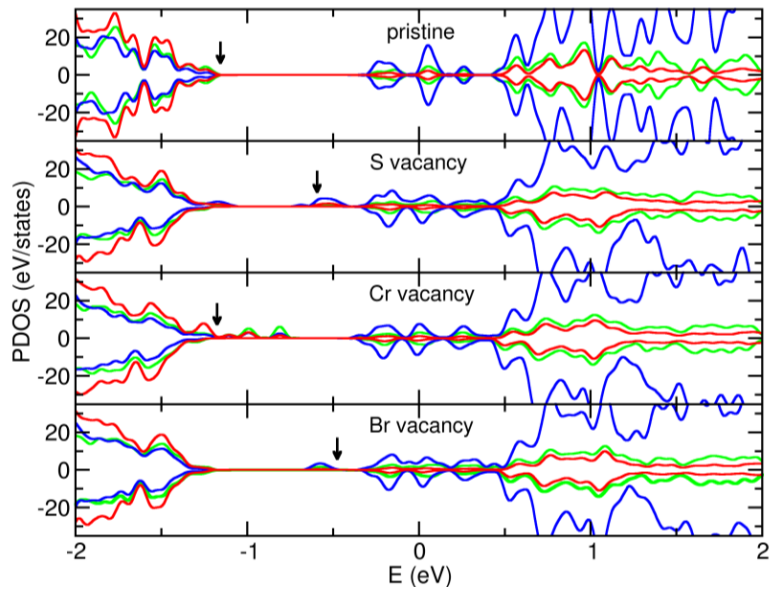

**Figure S6.** Projected density of states (PDOS) of pristine and defective CrSBr monolayers with S Cr, and Br vacancies. The blue, green, and red colors correspond to the projected states from the Cr, S, and Br atoms, respectively. The black arrow indicates the position of the highest occupied state.

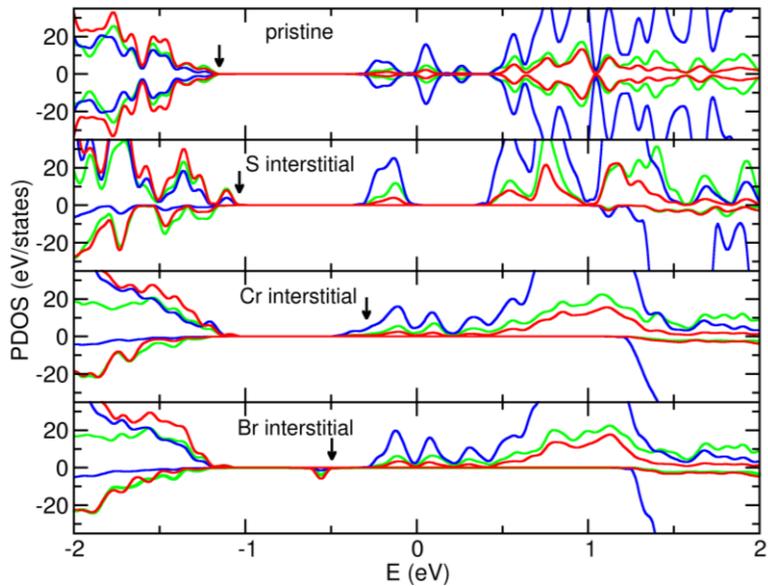

**Figure S7.** Projected density of states (PDOS) of pristine and defective CrSBr monolayers with S Cr, and Br interstitials. The blue, green, and red colors correspond to the projected states from the Cr, S and Br atoms, respectively. The black arrow indicates the position of the highest occupied state.

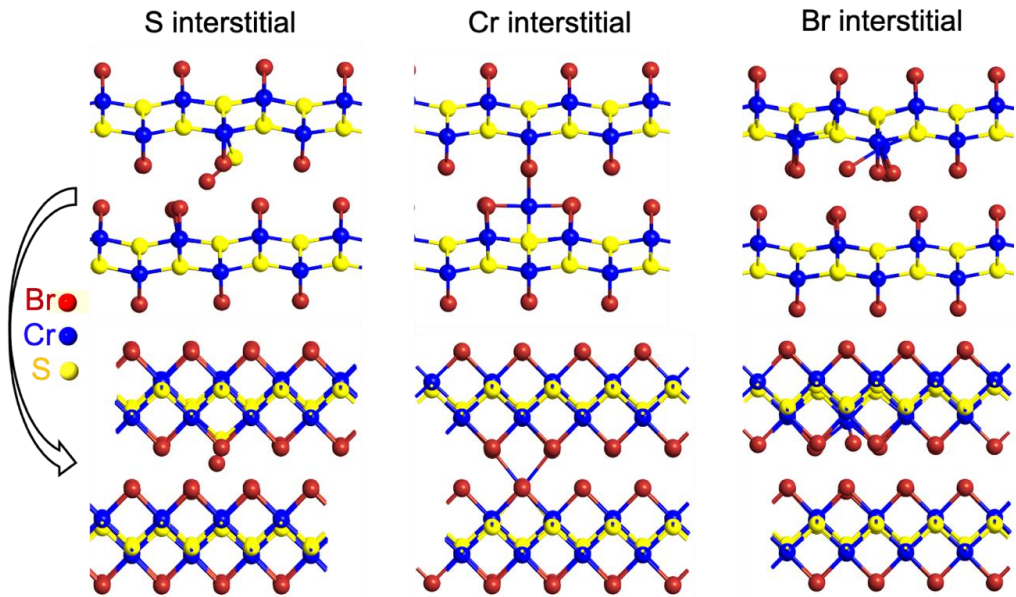

**Figure S8**. Optimized structures of CrSBr bilayer with different interstitials. Cr, S, and Br atoms are represented by blue, yellow, and red circles.

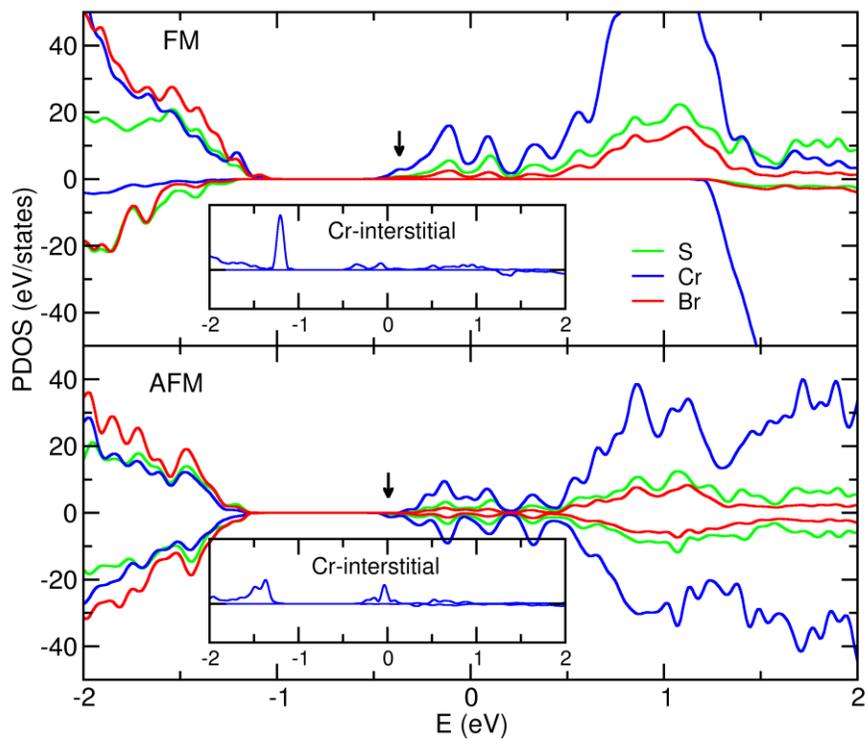

**Figure S9.** Projected density of states for CrSBr bilayer with Cr interstitial in FM and AFM ordering.

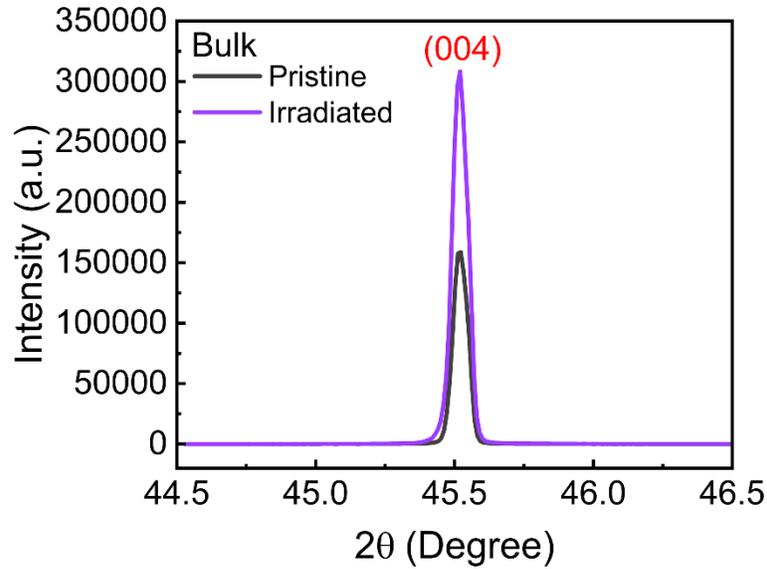

**Figure S10.** X-ray diffraction of pristine CrSBr and sample 8E15. By XRD, we cannot confirm any lattice expansion along the c-axis (the out-of-plane direction). It is much different in comparison with conventional semiconductor materials.[5] Here, only the CrSBr 004 reflection is shown since it is the strongest diffraction peak[6].

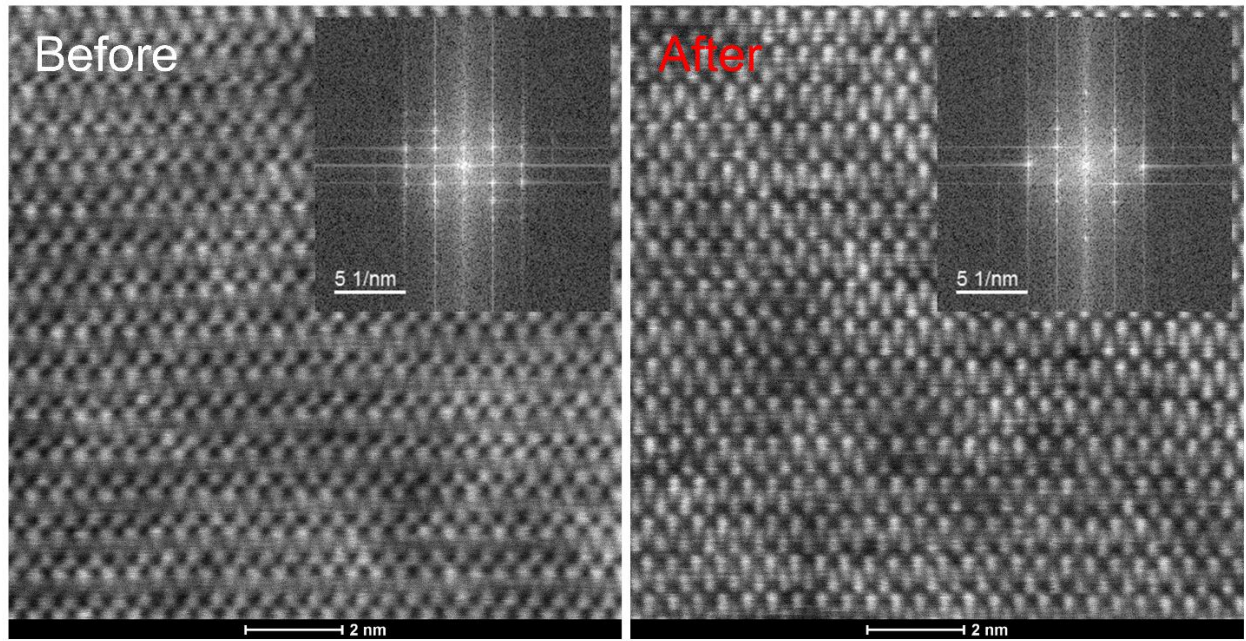

**Figure S11.** Cross-sectional high-resolution HAADF-STEM images and corresponding fast Fourier transforms (insets) of a new sample before (left) and after irradiation (right) with 4 MeV He ions at a fluence of $5\times10^{15}$/cm$^2$ which generates a similar concentration of defects (dpa) as for sample 4E15 ($4\times10^{15}$/cm$^2$) discussed in the main text. After irradiation, there are no structural differences visible compared to pristine CrSBr.

## Supporting references

(1) Kresse, G.; Furthmueller, J. Efficient iterative schemes for ab initio total-energy calculations using a plane-wave basis set. *Phys. Rev. B* **1996,** 54, 11169–11186.
(2) Kresse, G.; Furthmueller, J. Efficiency of Ab-Initio Total Energy Calculations for Metals and Semiconductors Using a Plane-Wave Basis Set. *Comput. Mater. Sci.* **1996,** 6 (27), 15–50.
(3) Perdew, J. P.; Burke, K.; Ernzerhof, M. Generalized Gradient Approximation Made Simple. *Phys. Rev. Lett.* **1996,** 77, 3865–3868.
(4) Grimme, S.; Antony, J.; Ehrlich, S.; Krieg, H. A consistent and accurate ab initio parametrization of density functional dispersion correction (DFT-D) for the 94 elements H-Pu. *J. Chem. Phys.* **2010,** 132 (15), 154104.
(5) Holland, O. W.; Budai, J. D.; White, C. W. Uniaxial lattice expansion of self ion implanted Si. *Appl. Phys. Lett.* **1990,** 57 (3), 243-245.
(6) Liu, W.; Guo, X.; Schwartz, J.; Xie, H.; Dhale, N. U.; Sung, S. H.; Kondusamy, A. L. N.; Wang, X.; Zhao, H.; Berman, D.; Hovden, R.; Zhao, L.; Lv, B. A three-stage magnetic phase transition revealed in ultrahigh-quality van der Waals magnet CrSBr. *ACS Nano* **2022,** 16, 15917-15926.